\documentclass{article}

\usepackage{arxiv}

\usepackage[utf8]{inputenc} 
\usepackage[T1]{fontenc}    
\usepackage{hyperref}       
\usepackage{url}            
\usepackage{booktabs}       
\usepackage{amsfonts}       
\usepackage{nicefrac}       
\usepackage{microtype}      
\usepackage{calc}
\usepackage{epsfig}
\usepackage{subcaption}
\usepackage{calc}
\usepackage{amssymb}
\usepackage{amstext}
\usepackage{amsmath}
\usepackage{amsthm}
\usepackage{multicol}
\usepackage{pslatex}
\usepackage{apalike,pifont}
\usepackage{graphicx}
\usepackage{doi}

\title{How to Model Privacy Threats \\ in the Automotive Domain}

\author{ \href{https://orcid.org/0000-0002-7045-0213}{\includegraphics[scale=0.06]{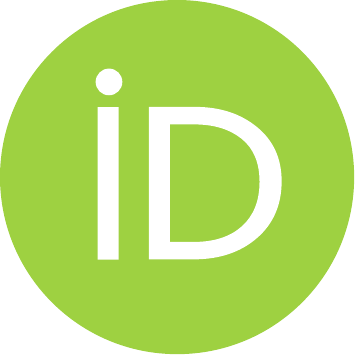}\hspace{1mm}Mario Raciti} \\
	IMT School for Advanced Studies Lucca\\
	Lucca, Italy \\
        Dipartimento di Matematica e Informatica\\
	Università di Catania\\
	Catania, Italy \\
	\texttt{mario.raciti@imtlucca.it} \\
	\And
	\href{https://orcid.org/0000-0002-7615-8643}{\includegraphics[scale=0.06]{orcid.pdf}\hspace{1mm}Giampaolo Bella} \\
	Dipartimento di Matematica e Informatica\\
	Università di Catania\\
	Catania, Italy \\
	\texttt{giamp@dmi.unict.it} \\
}

\renewcommand{\shorttitle}{How to Model Privacy Threats in the Automotive Domain}

\begin{document}
\maketitle

\begin{abstract}
This paper questions how to approach threat modelling in the automotive domain at both an abstract level that features no domain-specific entities such as the CAN bus and, separately, at a detailed level.
It addresses such questions by contributing a systematic method that is currently affected by the analyst's subjectivity because most of its inner operations are only defined informally. However, this potential limitation is overcome when candidate threats are identified and left to everyone's scrutiny.
The systematic method is demonstrated on the established LINDDUN threat modelling methodology with respect to 4 pivotal works on privacy threat modelling in automotive. As a result, 8 threats that the authors deem not representable in LINDDUN are identified and suggested as possible candidate extensions to LINDDUN. Also, 56 threats are identified providing a detailed, automotive-specific model of threats.
\end{abstract}


\section{Introduction}
\label{sec:introduction}

The world has realised that a whole new range of services tailored to car drivers' preferences and habits can be designed and made available, for example leveraging the computerised infrastructures of Smart Cars~\cite{enisa-report}, Smart Roads~\cite{POMPIGNA2022100986} and Smart Cities~\cite{smart-cities}.
Following a consolidated business model, such services can be delivered \textit{virtually for free} to drivers, namely at the sole price of enabling the service provider to act as a Data Controller or a Data Processor on behalf of each individual driver. In simpler terms, the price is to allow the service provider to treat the drivers' data according to the conditions specified in the driver's consent. Therefore, it is clear that privacy 
threats affect citizens also when they generate personal data by driving modern cars. 

The General Data Protection Regulation (GDPR) is the European answer to the privacy needs of its citizens, and is proving inspirational for other similar, international regulations. It warns about a \textit{personal data breach}, the ``\textit{accidental or unlawful destruction, loss, alteration, unauthorised disclosure of, or access to, personal data transmitted, stored or otherwise processed}'', which, therefore, may affect all scenarios in which \textit{personal data}, namely ``\textit{any information relating to an identified or identifiable natural person}'', are processed, which implies any form of ``\textit{collection, recording, organisation, structuring, storage, adaptation or alteration, retrieval, consultation, use, disclosure by transmission, dissemination or otherwise making available, alignment or combination, restriction, erasure or destruction}''~\cite{gdpr}.

The mentioned privacy issues in the automotive domain are perhaps insufficiently understood at present, but are certain to demand GDPR compliance. Compliance is meaningfully assessed in terms of privacy risk assessment, which in turn demands privacy threat modelling, hence the general motivation for this paper. Following the GDPR extracts given  above, a personal data breach represents the essential and most abstract version of a threat to citizens' privacy.
This is only an archetypal threat modelling exercise, while threat modelling is an established and challenging research area. 
%
%

\paragraph{Research questions and Contributions}
Threat modelling is challenging. The analyst faces a version of a soundness and completeness problem. Soundness may be interpreted at least as a sufficiently disambiguous and detailed description of each threat. However, completeness may be more impactful because failing to account for specific threats would cause pitfalls to the subsequent risk assessment. Completeness is also very challenging because the analyst needs to decide whether an extra threat is to be added to the current list, and this, in turn, raises two further problems. One is that the threat to be potentially added needs to be discovered. The other one is the scrutiny on whether that threat may be redundant with the current list, given that redundancy may lead to inconsistencies in the risk assessment. Even after appropriate decisions in this circumstance, the completeness problem reiterates. Therefore, the general research question that this paper sets for itself corresponds to its title: \textit{How to model privacy threats in the automotive domain?}

To go about such a question, we observe that a widely established privacy threat modelling methodology exists, LINDDUN~\cite{Deng2011}. So, an answer could be found, potentially, in such a methodology. However, LINDDUN is meant to be domain-independent, a feature that is bound to keep its threat descriptions only at an abstract level of detail. For example, threat $L\__{ds4}$ stands for ``Excessive data available''. A specific level of detail is not prescribed and depends on the analyst's knowledge and experience as applied to the specific exercise. 
%
%
However, we question whether LINDDUN is enough for the automotive domain, irrespectively of its level of detail, hence we set a specific research question: 
\begin{quote}
SRQ1. \textit{Can LINDDUN be considered complete, albeit abstract, when applied to automotive privacy?}
\end{quote}
Continuing our argument, it may be observed that the analyst may want a rather detailed model of threats for the automotive domain, namely a list of threats that 
specifically revolve around the typical entities involved in modern cars, such as threat ``CAN eavesdropping''.
This means that another specific research question arises:
\begin{quote}
SRQ2. \textit{Can we model detailed and specific privacy threats for the automotive domain that can be considered complete with respect to best practices from the state of the art?} 
\end{quote}

It is clear that answering the two specific research questions would answer the general research question. In other words, SRQ1 concerns whether LINDDUN suffices when threat modelling can be rather abstract, while SRQ2 concerns the same problem though at a detailed level, where no de facto standard methodology and only a few notable approaches exist, to the best of our knowledge.

This paper answers both the specific research questions by advancing a systematic method to completeness in threat modelling. Our method leverages LINDDUN, as it can be expected, and selects 4 relevant sources that are considered best practices in the state of the art:
\begin{enumerate}
    \item ``Good practices for security of smart cars''~\cite{enisa-report},
    \item ``Privacy threat analysis for connected and autonomous vehicles''~\cite{CHAH202236},
    \item ``A double assessment of privacy risks aboard top-selling cars''~\cite{Bella2023}.
    \item ``Calculation of the complete Privacy Risks list v2.0''~\cite{owasp},
\end{enumerate}
It can be seen that the final source pertains to web privacy, which can be justified on the basis of the tight interrelations with the automotive domain. 

Our systematic method counts 5 steps. Its gist is to derive a list of \textit{preliminary threats} from the 4 sources just itemised. Precisely, the preliminary threats are found to be 95. Then, these are polished according to various operations that we introduce below, to produce the \textit{final threats}. These amount precisely to 56 threats. We shall see that such final threats answer SRQ2. However, if we continue by mapping the final threats into LINDDUN threats, we find out that 8 threats could not be mapped, thereby concluding how LINDDUN could be potentially expanded, and effectively answering also SRQ1.

It is noteworthy that, although our independent modelling stems from the specific automotive application domain, the 8 threats that were left outstanding with respect to LINDDUN are general privacy threats, namely they ignore domain specific entities. This signifies that all domain specific threats could be mapped to more general LINDDUN threats.

\section{LINDDUN and Related Work}
\label{sec:background}
LINDDUN is a privacy threat modelling methodology, inspired by STRIDE, that supports analysts in the systematical elicitation and mitigation of privacy threats in software architectures. LINDDUN privacy knowledge support represents one of the main strength of this methodology, and it is structured according to the 7 privacy threat categories encapsulated within LINDDUN's acronym~\cite{Deng2011}, namely Linkability, Identifiability, Non-repudiation, Detectability, Disclosure of information, Unawareness, Non-compliance.

Landuyt et al.~\cite{landuyt2020} highlighted the influence of assumptions to the outcomes of the analysis during the risk assessment process, more precisely in the threat modelling phase in the context of a LINDDUN privacy threat elicitation.
Vasenev et al.~\cite{vehits19} were among the first to apply an extended version of STRIDE~\cite{stride} and LINDDUN~\cite{Deng2011} to conduct a threat analysis on security and privacy threats in the automotive domain. In particular, the case study is specific to long term support scenarios for over-the-air updates.
Chah et al.~\cite{CHAH202236} applied the LINDDUN methodology to elicit and analyse privacy requirements of CAV system, while respecting the privacy properties set by the GDPR. Such attempt represents a solid baseline for the broader process of privacy risk assessment tailored for the automotive domain.

\section{Our Systematic Method}
\label{sec:methodology}
Our method is systematic but not yet fully formalised. It means that the largest part of its steps and operations are still only informally defined, as already noted. This will become apparent below. We shall see that our findings are remarkable despite the currently mostly informal approach.

The pivotal notion that we rely upon is \textit{threat embracing}. It wants to capture the standard scrutiny that the analyst operates in front of a list of threats to understand the extent of their semantic similarity. One element of scrutiny here derives from the possible use of synonyms, for example a threat might mention the word ``protocol'' and a similar threat may just rewrite the first one by replacing that word with ``distributed algorithm''. Arguably, the analyst would conclude that these threats are embraceable and \textit{embrace them} by selecting the one with the wording that they find most appropriate, and discarding the other one.

Another element of scrutiny derives from the level of detail --- of the statement describing a threat. For example, ``Unchanged default password'' is certain to be more detailed than (the more abstract) threat ``Human error''. The analyst will typically conclude also in this case that these two threats are embraceable and proceed to embrace them by selecting the one whose level of detail they find most appropriate for the specific threat modelling exercise they are doing. Normally, the analyst strives to choose a consistent level of detail till the end of the exercise.

The five steps of our systematic method, supported by a running example, are detailed below.

\subsection{Step 1 --- Threat Collection}
\label{subsec:collection}
The first step involves the collection of the threats that the analyst deems relevant, namely arising from relevant sources. These may vary from case to case, and the analyst normally appeals to reliable academic publications, international standards, best-practice documents and other authoritative material from governmental bodies, research institutions and so on. In this paper, we selected the 4 relevant sources outlined above. There is no specific limit to the number of threats to be collected, and these may reach the order of hundreds, of course, depending once more on the application domain. There is also, in this step, no limit to the quality of threats that are collected, hence these are likely to bear redundancy and clear semantic similarity. These issues will be faced in the following steps.

In slightly more formal terms, this step is to build a list $P$ of \textit{preliminary threats} and assign an identifier to each threat so that:
\begin{tabbing}
\hspace{4mm}\= $P=p_1,\ldots,p_n$.
\end{tabbing} 
It is useful to organise the threats in a table, say ${\cal P}$, following the vertical dimension. Table ${\cal P}$ will grow with more and more columns as our systematic method proceeds. If $C_k$ is the projected function that takes a table and yields its $k$-th column, then:
\begin{tabbing}
\hspace{4mm}\= $C_1({{\cal P}})=P$.
\end{tabbing} 
The second column carries the threat description, or label in brief, for each threat by means of function $label$:
\begin{tabbing}
\hspace{4mm}\=
$C_2({\cal P})=LB$,\\[1ex]
\>where $LB = label(p_1),\ldots,label(p_n)$.
\end{tabbing}
We may also formalise the source where each threat derives from, which in this step corresponds to its origin document, by means of a function $source$. It is useful to note it down in the third column:
\begin{tabbing}
\hspace{4mm}\=
$C_3({\cal P})=S$,\\[1ex]
\>where $S = source(p_1),\ldots,source(p_n)$.
\end{tabbing}
A demonstration of this step on our running example yields a table with three columns that is omitted here but corresponds to the first three columns of Table~\ref{tab:re-step2}. 
Threat sources are only symbolically represented because the example is a mock-up.

\subsection{Step 2 --- Categorisation}
\label{subsec:categorisation}
The second step extends ${\cal P}$ by categorising each preliminary threat in $P$ with respect to the LINDDUN properties. In particular, we add a column to ${\cal P}$ for each of the seven properties, then we tick a cell if the given threat relates to that property. Obviously, a threat may apply to multiple LINDDUN properties. Such operations may be demanding because each depends on both the level of detail of the given threat and on the knowledge that is available on the target system and the threat scenarios. In particular, the categorisation step is prone to the analyst's bias subjectivity, as well as human errors. All this is further discussed below.

More formally, let us introduce boolean functions following the LINDDUN acronym $isL$, $isI$, $isN$, $isD$, $isDi$, $isU$ and $isNc$, which take a threat and hold when that threat applies to the respective property. For example:  

\[
    isN(t)= 
\begin{cases}
    \checkmark& \text{if the analyst decides that threat $t$ affects property $N$,} \\
    \text{\ding{55}}              & \text{otherwise}
\end{cases}
\]

In practice, the symbol \ding{55} is often omitted for readability, leaving an empty cell. Therefore, ${\cal P}$ grows as follows:

\begin{tabbing}
\hspace{4mm}\=
$C_4({\cal P})=L$,\ where $L = isL(p_1),\ldots,isL(p_n)$,\\[1ex]
\>$C_5({\cal P})=I$,\ where $I = isI(p_1),\ldots,isI(p_n)$,\\[1ex]
\>$C_6({\cal P})=N$,\ where $N = isN(p_1),\ldots,isN(p_n)$,\\[1ex]
\>$C_7({\cal P})=D$,\ where $D = isD(p_1),\ldots,isD(p_n)$,\\[1ex]
\>$C_8({\cal P})=Di$,\ where $Di = isDi(p_1),\ldots,isDi(p_n)$,\\[1ex]
\>$C_9({\cal P})=U$,\ where $U = isU(p_1),\ldots,isU(p_n)$,\\[1ex]
\>$C_{10}({\cal P})=Nc$,\ where $Nc = isNc(p_1),\ldots,isNc(p_n)$.
\end{tabbing}
Table~\ref{tab:re-step2} shows the threats put together through the previous step now enriched with appropriate ticks to highlight the concerned LINDDUN properties. In particular, it can be noticed that all threats concern linkability.

\begin{table*}[ht]
\caption{Outcomes of Step 1 and Step 2 on our running example.}\label{tab:re-step2} \centering
\begin{tabular}{|c|c|c|c|c|c|c|c|c|c|}
  \hline
  $P$ & $LB$ & $S$ & $L$ & $I$ & $N$ & $D$ & $Di$ & $U$ & $Nc$ \\
  \hline
  $p_1$ & Insufficient randomness of session ID & $source(p_1)$ & \checkmark &  &  &  &  &  & \\
  \hline
  $p_2$ & Session control mechanisms may be hijacked & $source(p_2)$ & \checkmark &  &  &  &  & & \\
  \hline
  $p_3$ & Browser is not updated & $source(p_3)$ & \checkmark & \checkmark &  &  & \checkmark &  & \\
  \hline
\end{tabular}
\end{table*}

\subsection{Step 3 --- Manipulation}
\label{subsec:manipulation}
The third step shapes a new list $F$ of threats that we call \textit{final threats} and store them in the first column of a new table $\cal F$. Formally:
\begin{tabbing}
\hspace{4mm}\= $F=f_1,\ldots,f_m$,\\[1ex]
\hspace{4mm}\= $C_1({\cal F})=F$.
\end{tabbing} 
Columns from 4 through to 10 in ${\cal F}$ are defined as with ${\cal P}$ but over $F$ rather than over $P$. We need to specify how to fill the new table up.
The underlying concept is to build this table to solve the redundancies arisen in the old table. In fact, the use of different sources inherently leads to different levels of detail and various overlaps, beside the fact that some entries could refer to the same threat scenario. To address these issues, we define a list of operations to build the final threats upon the basis of the preliminary threats. The range of the various indexes is self-evident and omitted here for brevity.

The first operation applies to a list of generic threats $t_1,\ldots,t_s$ (namely, they could be either preliminary or final) which are considered embraceable:

\begin{tabbing}
\hspace{4mm}\= $o_1.\ \mathit{embrace}(t_1,\ldots,t_s)$. 
\end{tabbing}
The result of this operation is a threat that gets the same id and label as the element of the input threat with the most pertaining level of detail according to the analyst. Otherwise, if all preliminary threats that are considered bear a similar level of detail, then the final threat gets the same label as the first element of the list. The LINDDUN properties corresponding to the computed threat are the union of all properties that were ticked for each of the input threats $t_1,\ldots,t_s$. In general, threats can be embraced together multiple times, both at preliminary and at final levels. This operation is useful to build $\cal F$, hence to build a final threat from given preliminary threats, namely:
\begin{tabbing}
\hspace{4mm}\= $f_l := embrace(p_i, \dots, p_j)$.
\end{tabbing}
The second operation renames a threat:
\begin{tabbing}
\hspace{4mm}\= $o_2.\ \mathit{rename}(t_q)$.
\end{tabbing}
The analyst may judge the default label as incomplete and feel the need to modify the level of detail of the threat label, while the ticked LINDDUN properties remain unvaried. This is specifically useful, for example, when we want to assign a proper label to a threat in $\cal F$ produced by an embrace of threats:
\begin{tabbing}
\hspace{4mm}\= $f := rename(f)$.
\end{tabbing}
The last operation discards a threat, meaning that the considered threat is excluded from the current table (and possibly moved to a reserve list for future re-inspection):
\begin{tabbing}
\hspace{4mm}\= $o_3.\ \mathit{discard}(t)$.
\end{tabbing}
This is necessary when a threat is inapplicable to the domain, e.g., it strictly refers to security rather than privacy or is considered irrelevant for the particular target system. An example application is to a preliminary threat, which is therefore not going to be reported in $\cal F$. This is the only operation that can be defined formally here. If index 0 denotes an empty threat, we have that:
\begin{tabbing}
\hspace{4mm}\= $discard(t) = t_0$.
\end{tabbing}
It is crucial to apply the operations above with caution to avoid the loss of relevant information and maintain the semantic of the threats unvaried.
Moreover, operations can be nested. For example, given a list of threats referring to the same threat scenario, it may happen that none of them embraces the others, thus the resulting label of an embrace operation would be inappropriate. This issue can be addressed by nesting the first two operations as follows: $rename(embrace(p_i, \dots, p_j))$.

On our running example, we observe that $p_1$ and $p_2$ are embraceable, hence we apply the $embrace$ operation. We consider $p_2$ more general than $p_1$, hence we set:
\begin{tabbing}
\hspace{4mm}\= $embrace(p_1, p_2)=p_2$,\\[1ex]
\>$f_1 := p_2$.
\end{tabbing}
We then rename the outcome by operation $rename$:
\begin{tabbing}
\hspace{4mm}\= $rename(p_2)=$ ``Weak web session control mechanisms''.
\end{tabbing}
Finally, we observe that preliminary threat $p_3$ is a verifiable event, namely its likelihood would be null or top in a given scale. We decide that this event belongs more correctly to the list of security measures that can be verified by controls, rather than to a threat list. Therefore, we apply $discard(p_3)$. The final outcome of this list of operations is shown in Table~\ref{tab:re-step3}.

\begin{table*}[ht]
\caption{Outcomes of Step 3 on our running example.}\label{tab:re-step3} \centering
\begin{tabular}{|c|c|c|c|c|c|c|c|c|c|}
  \hline
  $F$ & $LB$ & $S$ & $L$ & $I$ & $N$ & $D$ & $Di$ & $U$ & $Nc$ \\
  \hline
  $f_1$ & Weak web session control mechanisms
 & $rename(embrace(p_1, p_2))$ & \checkmark & &  &  &  & & \\
  \hline
\end{tabular}
\end{table*}

Having reached the end of this step, table $\cal F$ of final threats can be leveraged to answer SRQ2, as we shall see in Section~\ref{sec:case-studies}.

\subsection{Step 4 --- Mapping}
\label{subsec:mapping}
The fourth step consists in verifying whether the threat catalogue proposed by the LINDDUN framework covers the threats in $\cal F$ and vice versa. This can be done by appropriate applications of the $embrace$ operations, as detailed here. 

For each final threat $f$ and each of the properties that are ticked in $\cal F$ for it, we study the corresponding LINDDUN property tree to distil out all nodes that are  embraceable with $f$ and then apply the operation. The analyst should proceed carefully to make sure that every $embrace$ operation yields a LINDDUN threat because this is useful to address the research questions stated above. By contrast, such a requirement through the application of the operation could be removed should the analyst have a different aim, for example of modelling what they find the best threats according to their own knowledge and experience.

By proceeding systematically, a new table $\cal M$ can be built, representing all \textit{mapped threats}, namely all LINDDUN threats that could embrace a final threat. Also, this table has the usual structure, but columns numbered 4 through to 10 can be omitted because only LINDDUN threats are represented and their identifier suggests the overarching property.
For example, the LINDDUN nodes in the Linkability property tree that we deem embraceable with $f_1$ are:
\begin{tabbing}
\hspace{4mm}\= $L\__{df1}=$ ``Linkability of transactional data (transmitted data)'',\\[1ex]
\> $L\__{df4}=$ ``Non-anonymous communication are linked'',\\[1ex]
\> $L\__{df10}=$ ``Based on session ID''.
\end{tabbing}
Therefore, we calculate:
\begin{tabbing}
\hspace{4mm}\= $embrace(f_1, L\__{df1}, L\__{df4}, L\__{df10})=L\__{df10}$
\end{tabbing}
and assign:
\begin{tabbing}
\hspace{4mm}\= $m_1 := L\__{df10}$
\end{tabbing}
 
The operation yields $L\__{df10}$, which, according to the LINDDUN notation, must be read as ``Non-anonymous communication are linked based on session ID''. This particular embrace is coherent with the aim to answer SRQ1, as we shall see in Section~\ref{sec:case-studies}.
Of course, if other LINDDUN properties were ticked, the list of threats to embrace should include the additional LINDDUN threats that would be embraceable with $f_1$, taken from the corresponding property trees.

It may now be the case that the
analyst does not feel like mapping some final threats to any of the LINDDUN ones. It means that the analyst feels that no LINDDUN trace is embraceable with those final threats. When this is the case, our systematic method would highlight a limitation of LINDDUN in terms of coverage. By taking typical domain details off the final threats that could not be mapped, we get a list of valid candidates to become new nodes in the pertaining threat tree(s) of an amended LINDDUN methodology.

Finally, it is noteworthy that this step implicitly provides the opportunity to adjust potential errors from Step 2 as it provides a more granular view thanks to the significant amount of nodes to examine. 

\subsection{Step 5 --- Safety Check}
\label{subsec:safety-check}
The last step implements a further safety check of Step 2, when we may have assigned an insufficient list of pertaining properties to the preliminary threats that were later embraced in some final threat. To thwart that, this step prescribes, for each final threat $f$, the analyst to assess all LINDDUN property trees as it was done in Step 4 for the pertaining properties only. The clear aim is to find any LINDDUN threat at all that would be embraceable with $f$.

Furthermore, this step is relevant because the assignment of properties to threat was only done with preliminary threats. The final threats may include, for example after the analyst's renaming operation, a level of detail that may highlight some link with the LINDDUN properties. Therefore, this step is crucial to also minimise the odds of erroneous exclusions, which would lead the analyst to conclude that certain final threats could not be mapped into LINDDUN.

For example, following in-depth scrutiny, we may now observe that $f_1$ also concerns Identifiability and Disclosure of information properties, due to threats: 
\begin{tabbing}
\hspace{4mm}\= $I\__{df1}=$ ``Identifiability of transactional data (transmitted data)'',\\[1ex]
\> $I\__{df6}=$ ``Non-anonymous communication traced to entity'',\\[1ex]
\> $I\__{ds2}=$ ``Non-anonymous communication are linked'',\\[1ex]
\> $I\__{df10}=$ ``Based on session ID''.
\end{tabbing}
Therefore, we update $m_1$ by means of a further embrace operation that is larger than the previous one:
\begin{tabbing}
\hspace{4mm}\= $m_1 := embrace($\=$f_1, I\__{df1}, I\__{df4}, I\__{df10}, I\__{df1},I\__{df6},I\__{ds2},I\__{df10})$.
\end{tabbing}
This means that the analyst gains an additional opportunity to decide how to best represent $f_1$ within LINDDUN. 

\section{Demonstration of our Method}
\label{sec:case-studies}
We apply our systematic method described above to address the specific research questions.
The full outcomes, including the 95 preliminary and the 56 detailed, final privacy threats for the automotive, are released on a GitHub repository~\cite{repo}. In particular, the latter threats, built by taking our systematic method up to Step 3, answer SRQ2. We provide two distinct Excel files, reflecting the three automotive sources and the web application source separately. Each file contains sheets named according to the same terminology introduced in Section~\ref{sec:methodology}, namely results from Step 1 are included in the sheet "Step 1", and so on.

\subsection{The 3 Sources from Automotive}
\label{subsec:automotive-results}
The following paragraphs reflect the steps of our systematic method and feature a few notable examples for the sake of brevity.

\paragraph{Collection}
We selected three sources of threats that pertain to the automotive domain. To do so, we appealed to a best-practice document, namely Good practices for security of Smart Cars report~\cite{enisa-report}, and two recent and reliable academic publications~\cite{CHAH202236}~\cite{Bella2023}. The report proposed by ENISA provides a list of relevant threats and risks with a focus on “cybersecurity for safety”. The second contribute~\cite{CHAH202236} provides an extract of some vulnerabilities and privacy-related attack scenarios onboard and outboard connected and autonomous vehicles (CAVs). Furthermore, the third and last source~\cite{Bella2023} features a list of privacy threats targeting the automotive domain.
Following our systematic method, we collected a total of 75 preliminary threats, distributed as Table~\ref{tab:step-1} shows.

\begin{table}[h]
\caption{Distribution of preliminary threats over the 3 automotive privacy sources}\label{tab:step-1} \centering
\begin{tabular}{|c|c|}
  \hline
  Source & Number of threats \\
  \hline
  ENISA & 30 \\
  \hline
  Chah et al. & 20 \\
  \hline
  Bella et al. & 25 \\
  \hline
  Total & 75 \\
  \hline
\end{tabular}
\end{table}

\paragraph{Categorisation}
Successively, we applied a categorisation of the 75 threats. In particular, the first source (ENISA) provides a threat taxonomy that includes descriptions of the threats. Therefore, we leveraged such descriptions to better identify the LINDDUN properties affected by those threats. Furthermore, the second source~\cite{CHAH202236} offers a view of the privacy threats along with the attack scenarios, preconditions and the LINDDUN properties affected. We trust the work of the authors, thus for each threat from this source we crossed the very same LINDDUN properties. The third source~\cite{Bella2023} derives the list of privacy threats from a STRIDE threat modelling and justifies them in prose, thereby we leveraged such descriptions to identify, once more, the affected LINDDUN properties.

\paragraph{Manipulation}
At this point, we expected the three different sources to provide threat labels with various levels of detail and different terminology. Therefore, we adopted all the three operations discussed above for this step to slim down the list of preliminary threats. For the sake of simplicity, Table~\ref{tab:step-3a} shows an extract of the final threats as a derivation process for three illustrative threats, namely $f_{21}$ ``Infotainment alteration'', $f_{35}$ ``Unauthorised access \textit{in OEM and/or car services}'' and $p_{30}$ ``Car depleted battery''. 

In detail, $f_{21}$ ``Infotainment alteration'' is derived by $embrace$ over the following threats: $p_{73}$ ``Infotainment alteration'', $p_{3}$ ``Manipulation of hardware and software'', $p_{37}$ ``An adversary can execute arbitrary code on the telematics unit (TCU) and take control of the device.'', $p_{43}$ ``Attacker operates physically on the TC by tampering the device firmware.'' and $p_{44}$ ``An attacker could perform remote control by installing remotely his own software on the device.'' 

By contrast, $f_{35}$ ``Unauthorised access \textit{in OEM and/or car services}'' is obtained via a combination of  operations $rename$ and $embrace$ of the threats: $p_{63}$ ``Unauthorised diagnostic access'', $P_{11}$ ``Unauthorised activities''. Finally, $p_{30}$ ``Car depleted battery'' is discarded thanks by $discard$, since we deemed it irrelevant as a privacy threat.

\begin{table}[h]
\caption{Example of threat finalisation.}\label{tab:step-3a} \centering
\begin{tabular}{|c|c|c|}
  \hline
  $F$ & $S$ \\
  \hline
  $f_{21}$ & $embrace(p_{73}, p_3, p_{37}, p_{43}, p_{44})$\\
  \hline
  $f_{35}$ & $rename(embrace(p_{11}, p_{63}))$ \\
  \hline
  $f_0$ & $discard(p_{30})$ \\
  \hline
\end{tabular}
\end{table}

We end up with 41 final threats, as Table~\ref{tab:step-3b} illustrates along with some statistics on the number of operations applied.

\begin{table}[h]
\caption{Operations applied in the automotive domain.}\label{tab:step-3b} \centering
\begin{tabular}{|c||cccc|}
\hline
{Step 2} & \multicolumn{4}{c|}{{Step 3}} \\
\hline
Total & \multicolumn{1}{c||}{Total} & \multicolumn{1}{c|}{Embrace} & \multicolumn{1}{c|}{Rename} & Discard \\
\hline
75 & \multicolumn{1}{c||}{41} & \multicolumn{1}{c|}{26} & \multicolumn{1}{c|}{4} & 3 \\
\hline
\end{tabular}
\end{table}

\paragraph{Mapping}
For the sake of simplicity, we present an extract of the mapping with respect to the threat tree for a LINDDUN property, precisely the Detectability property. 
The leading final threats here are $f_7$ ``Communication protocol hijacking \textit{in car devices}'', $f_{32}$ ``Software vulnerabilities exploitation \textit{in OEM and/or car services}'' and $f_{35}$ ``Unauthorised access \textit{in OEM and/or car services}''. In particular, we performed the following operations:

\begin{tabbing}
\hspace{4mm}\= $embrace(f_{32}, f_{35}, D\__{ds1})=D\__{ds1}$,\\
\hspace{4mm}\= $embrace(f_{32}, D\__{ds2})=D\__{ds2}$,\\
\hspace{4mm}\= $embrace(f_{7}, D\__{ds3})=D\__{ds3}$.
\end{tabbing}


We could not match the following threats with any LINDDUN node: $f_{13}$ ``Failure to meet contractual requirements \textit{with driver}'' and $f_{41}$ ``Violation of rules and regulations/Breach of legislation/ Abuse of \textit{driver} personal data''.

\paragraph{Safety Check}
Finally, we iterated over all the nodes of the trees, independently of the LINDDUN property, but did not find any additional LINDDUN threats to which $f_{13}$ and $f_{41}$ could be reasonably mapped.

\subsection{The Source from the Web Domain}
\label{subsec:general-results}
This Section discusses specificities arisen from the OWASP list of threats for web privacy, which we find relevant for the wider automotive domain.

\paragraph{Collection}
We considered a general list of privacy threats targeting web applications, namely OWASP Top 10 Privacy Risks\cite{owasp}, as the unique source during collection due to its relevance in the chosen domain.
In particular, we employed the ``Calculation of the complete Privacy Risks list v2.0'', which includes a total of 20 threats forming the preliminary threats according to our systematic method.

\paragraph{Manipulation}
Successively, we realised that some operations were needed to overcome  redundancy, hence we went through various applications of $embrace$ and $rename$ but never used $discard$. The preliminary threats reduced to a total of 15 final threats, and some relevant statistics are in Table \ref{tab:step-3o}. As a result, we could not match the following threats with any LINDDUN nodes: $f_{2}$ ``Consent-related issues \textit{with driver}'', $f_{4}$ ``Inability of \textit{driver} to access and modify data'', $f_{7}$ ``Insufficient data breach response \textit{from OEM}'', $f_{11}$ ``Misleading content \textit{in OEM services}'', $f_{13}$ ``Secondary use \textit{of driver data}'' and $f_{14}$ ``Sharing, transfer or processing through 3rd party \textit{of driver data}''.

\begin{table}[h]
\caption{Operations applied in the web domain.}\label{tab:step-3o} \centering
\begin{tabular}{|c||cccc|}
\hline
{Step 2} & \multicolumn{4}{c|}{{Step 3}} \\
\hline
Total & \multicolumn{1}{c||}{Total} & \multicolumn{1}{c|}{Embrace} & \multicolumn{1}{c|}{Rename} & Discard \\
\hline
20 & \multicolumn{1}{c||}{15} & \multicolumn{1}{c|}{3} & \multicolumn{1}{c|}{4} & 0 \\
\hline
\end{tabular}
\end{table}

\begin{table*}[ht]
\caption{Final threats from the automotive and web domains that we could not match to any LINDDUN threat.}\label{tab:results} \centering
\begin{tabular}{|l|l|l|}
\hline
$F$       & $LB$                                                                                                                       & $S$                               \\ \hline
$f_{13a}$ & Failure to meet contractual requirements \textit{with driver}                                                                                  & $p_{27a}$                          \\ \hline
$f_{41a}$ & \begin{tabular}[c]{@{}l@{}}Violation of rules and regulations/Breach of legislation/\\ Abuse of \textit{driver} personal data\end{tabular} & $p_{28a}$                          \\ \hline
$f_{2w}$  & Consent-related issues \textit{with driver}                                                                                                    & $rename(embrace(p_{4w}, p_{17w}))$    \\ \hline
$f_{4w}$  & Inability of \textit{driver} to access and modify data                                                                               & $p_{9w}$                             \\ \hline
$f_{7w}$  & Insufficient data breach response \textit{from OEM}                                                                                         & $p_{3w}$                             \\ \hline
$f_{11w}$ & Misleading content \textit{in OEM services}                                                                                                        & $p_{16w}$                          \\ \hline
$f_{13w}$ & Secondary use \textit{of driver data}                                                                                                             & $p_{19w}$                          \\ \hline
$f_{14w}$ & Sharing, transfer or processing through 3rd party \textit{of driver data}                                                                         & $rename(embrace(p_{12w}, p_{15w}))$ \\ \hline
\end{tabular}
\end{table*}

\subsection{Findings and Conclusions}
\label{subsec:findings}
The full final threats are available online~\cite{repo}.
The application of our systematic method highlighted that there are final threats that are not embraceable with any LINDDUN node according to the analyst's judgement, and these are summarised in Table~\ref{tab:results}. Note that the table relies on a suffix to the indexes of the threats to avoid ambiguity, namely threats from Section~\ref{subsec:automotive-results} are referred to as $f_ia$, whilst those from Section~\ref{subsec:general-results} are indicated as $f_iw$, to distinguish the a(utomotive) domain from the w(eb application) one. By taking off the phrases in italics, we get a list of threats that are general enough to become valid candidates as new nodes in the pertaining threat tree(s) of an amended LINDDUN methodology. This answers SRQ1, which required the execution of our systematic method up to its final step.


This paper faced the challenge of threat modelling in the automotive domain in two ways. It questioned whether LINDDUN could suffice as an abstract-level methodology, concluding that it may have to be extended with 8 new threats, thereby effectively answering SRQ1. It questioned how to build a list of detailed threats in the same domain ensuring that the list is complete with respect to chosen relevant best practices, concluding with a list of 56 detailed, final threats, thereby effectively answering SRQ2.


The paper has remarked consistently that its findings are biased by the authors' subjectivity. However, all identified threats remain valid candidates for the international community's evaluation. 
While it seems a stretch to imagine that the analyst's role may be emptied entirely, our future research looks at modern, intelligent techniques from the area of Natural Language Processing to improve the formalisation of the various operations made through the steps of our systematic method. In particular, the upcoming steps involve the application of Semantic Similarity to score the relationship between threats based on their semantic, hence ultimately reducing subjectivity. 


\bibliographystyle{unsrtnat}
\bibliography{references}  






\end{document}